\documentclass[aps,pra,twocolumn,superscriptaddress,showpacs]{revtex4}
\usepackage{graphicx,latexsym}

\begin{document}

\title{Dynamical creation of entanglement by homodyne-mediated feedback}

\author{Jin Wang}
\affiliation{Department of Physics and Astronomy, The University of Nebraska, Lincoln, Nebraska 68588-0111,USA}
\affiliation{Centre for Quantum Computer Technology, The University of
Queensland, Brisbane, Queensland 4072, Australia}

\author{H. M. Wiseman}
\address{Centre for Quantum Computer Technology, Centre for Quantum Dynamics, School of Science, Griffith University, Brisbane, Queensland 4111, Australia}

\author{G. J. Milburn}
\affiliation{Centre for Quantum Computer Technology, The University of
Queensland, Brisbane, Queensland 4072, Australia}

\date{1st July 2004}

\begin{abstract}
For two two-level atoms coupled to a single-mode cavity field that
is driven and heavily damped, the steady-state can be entangled by
shining an un-modulated driving laser on the system [S.Schneider,
G. J. Milburn Phys. Rev A {\bf 65}, 042107, 2002]. We present a
scheme to significantly increase the steady-state entanglement by
using homodyne-mediated feedback, in which the driving laser is
modulated by the homodyne photocurrent derived from the cavity
output. Such feedback can increase the nonlinear response to both
the decoherence process of the two-qubit system and the coherent
evolution of individual qubits. We present the properties of the
entangled states using the SO(3) $Q$ function.
\end{abstract}


\maketitle

\newcommand{\beq}{\begin{equation}}
\newcommand{\eeq}{\end{equation}}
\newcommand{\bqa}{\begin{eqnarray}}
\newcommand{\eqa}{\end{eqnarray}}
\newcommand{\nn}{\nonumber}
\newcommand{\nl}[1]{\nn \\ && {#1}\,}
\newcommand{\erf}[1]{Eq.~(\ref{#1})}
\newcommand{\rf}[1]{(\ref{#1})}
\newcommand{\dg}{^\dagger}
\newcommand{\rt}[1]{\sqrt{#1}\,}
\newcommand{\smallfrac}[2]{\mbox{$\frac{#1}{#2}$}}
\newcommand{\half}{\smallfrac{1}{2}}
\newcommand{\bra}[1]{\langle{#1}|}
\newcommand{\ket}[1]{|{#1}\rangle}
\newcommand{\ip}[2]{\langle{#1}|{#2}\rangle}
\newcommand{\sch}{Schr\"odinger }
\newcommand{\schs}{Schr\"odinger's }
\newcommand{\hei}{Heisenberg }
\newcommand{\heis}{Heisenberg's }
\newcommand{\bl}{{\bigl(}}
\newcommand{\br}{{\bigr)}}
\newcommand{\ito}{It\^o }
\newcommand{\str}{Stratonovich }
\newcommand{\dbd}[1]{\frac{\partial}{\partial {#1}}}
\newcommand{\sq}[1]{\left[ {#1} \right]}
\newcommand{\cu}[1]{\left\{ {#1} \right\}}
\newcommand{\ro}[1]{\left( {#1} \right)}
\newcommand{\an}[1]{\left\langle{#1}\right\rangle}
\newcommand{\implies}{\Longrightarrow}


\section{Introduction}

The deeper ways that quantum information differs from classical
information involve the properties, implications, and use of
quantum entanglement. Entangled states are interesting because
they exhibit correlations that have no classical analog. Any two
systems described by a pure state that cannot be expressed as a
direct product $\ket{\Psi}=\ket{A}{\otimes}\ket{B}$ are entangled
(non-separable). For a two-qubit system there are four mutually
orthogonal Bell states, which may be denoted \bqa
\ket{\phi^{\pm}}_{AB}&=&\frac{1}{\sqrt{2}}\ro{\ket{00}_{AB}\pm\ket{11}_{AB}},\nn\\
\ket{\psi^{\pm}}_{AB}&=&\frac{1}{\sqrt{2}}\ro{\ket{01}_{AB}\pm\ket{10}_{AB}}
\eqa where $A$ and $B$ are two subsystems, $\ket{0}$ and $\ket{1}$
are two orthogonal states that could represent the atomic ground
and excited state, respectively.

Quantum entanglement is a subject of intensive study because it is
useful and frequently essential for quantum
teleportation~\cite{Bennett,Vitali,Werner}, quantum
cryptography~\cite{shor,Buttler}, quantum dense
coding~\cite{Werner} and quantum computation~\cite{Lo,munro}.
Also, entangled atoms can be used to improve frequency standards
{~\cite {Dowling1}}. Thus, one of the increasing interests in this
context is to find ways to generate the right type of entanglement
as well as possible~\cite{Vedral}. There are a number of measures,
like the concurrence \cite{Wootters},
 entanglement of formation \cite{Wootters}, entanglement of distillation
 \cite{Rains,Vedral}, relative entropy of
 entanglement \cite{Neumann}, and negativity \cite{Eisert,zyczkowski}, that have been
 proposed in recent years for the purpose of quantifying the amount of entanglement.

In this paper we address the question: what is the maximum amount
of steady-state entanglement that can be generated in a system
consisting of two two-level atoms (qubits) collectively damped, with the
output being measured and fed-back to control the system state?
This question follows naturally from the study of Schneider and Milburn
\cite{thesis1} which considered the same system but without measurement
or feedback. In both cases the entangled steady state is a
mixed state while most quantum entanglement algorithms are
designed for ideal pure states. However, it is usually very hard
to create, maintain, and manipulate pure entangled states under
realistic conditions, simply because any system is subject to the
interactions with its external environment. These effects, having
their origin in decoherence, may turn pure state entanglement into
mixed state entanglement. Therefore it immediately raises the
question of whether the entanglement can be distilled and used as
a resource for some quantum communication or computation task.

The paper is organized as follows. The model and the corresponding
master equation together with the derivation of the adiabatic
elimination of the cavity mode for a two-qubit system are analyzed
and the steady-state solution is presented in Section {\ref
{model}}. The entanglement measures and the relationship between
purity and entanglement are analyzed in Section {\ref {measure}}.
In section {\ref {Q and}}, we investigate the behaviour of the $Q$
distribution function and the density matrix in order to obtain
information about the entangled states. We discuss our results in
the concluding section {\ref {discussion}}.

\section{Model and master equation}\label{model}
A quantum computer requires that a set of $N$ two-state systems
can be prepared in an arbitrary superposition state. Each of these
systems is said to encode a qubit, as distinct from the single bit
encoded in a classical two state system.

We consider the case where two qubits are coupled to a single
cavity mode that is driven and heavily damped. The two qubits
undergo spontaneous emission into the cavity mode, and the
fluorescence from the output of the cavity is subjected to a
homodyne measurement with a subsequent feedback to the cavity. Our
primary goal of feedback control in this paper is to demonstrate
the ability of feedback control to increase steady-state
entanglement by counteracting the effects of both spontaneous
emission and the measurement back-action on the system. The
history of feedback control in open quantum systems goes back to
the 1980s with the work of Yamamoto and co-workers
\cite{Yamamoto}, and Shapiro and co-workers\cite{Shapiro}. Their
objective was to explain the observation of fluctuations in a
closed-loop photocurrent. They did this using quantum Langevin
equations (stochastic Heisenberg equations for the system
operators) and also semiclassical techniques. The latter approach
was made fully quantum-mechanical by Plimak. For systems with
linear dynamics, all of these approaches, and the quantum
trajectory approach of Refs.\cite{Car93b,WisMil93a}, are
equally easy to use to find analytical solutions. The advantage of
Wiseman and Milburn's (quantum trajectory) approach to quantum
control via feedback is for systems with nonlinear dynamics, as we
will discuss in this paper.

To obtain a master equation for the two-qubit system, we follow
the analysis of Schneider and Milburn's
 paper \cite {thesis1} for the steady-state of a system \cite{Cakir} of two qubits interacting simultaneously
with a driving laser. They have shown that the steady state is
entangled. In this work, we expand their analysis to include
feedback modulation of the amplitude of the driving laser.

\begin{figure}
\centerline{\scalebox{.4}{\includegraphics{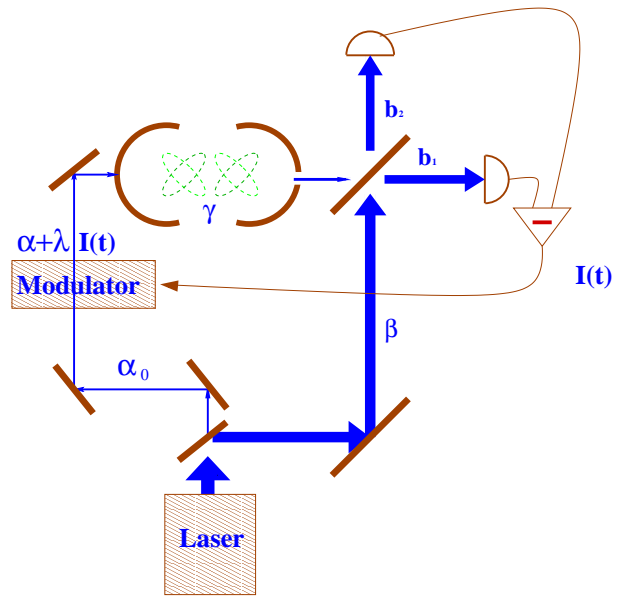}}}
\caption[Feedback diagram for dynamical creation of steady-state
entanglement for two qubits.]{Diagram of the experimental
apparatus. The laser beam is split to produce both the local
oscillator $\beta$ and the field $\alpha_{0}$ which is modulated
using the homodyne current $I(t)$ derived from the damped
cavity.}\label{diag22}
\end{figure}

A schematic diagram of the apparatus is shown in Fig.\ref{diag22}.
To ensure that the two atoms see the same phase and amplitude of
the cavity mode, we need to locate them at points of the standing
wave in which the two atoms are separated by an integer number of
wavelengths. The dipole-dipole interaction between the atoms can
thus be ignored in the following calculation. We assume a strong
coupling $g$ between the atoms and the cavity mode. The fact that
the coupling is the same for both atoms means that they are
indistinguishable. This leads to interference in their damping via
the cavity, as we shall see. Such a proposed approach can be
realized experimentally in a cavity QED system. In this paper we
reproduce the results in \cite {thesis1} and show that one can
increase the steady-state entanglement by using
feedback-modulation of the laser that drives the cavity mode.

It is assumed that that the laser interacts with the two atoms
simultaneously, forcing each atom to undergo Rabi oscillations at
the same frequency and phase. We can thus define the total angular
momentum operators for the two qubits:
 \bqa
\hat{J}^{\pm}&=&\hat{\sigma}^{\pm}_1+\hat{\sigma}^{\pm}_2. \\
&=&\hat{J}_{x}\pm{i}\hat{J}_{y},\eqa
where \beq
\hat{\sigma}^{\pm}=\hat{\sigma}_{x}\pm{i}\hat{\sigma}_{y}\eeq
In terms of these operators, the cavity-atom interaction
Hamiltonian is: \bqa
\hat{H}=\frac{g}{2}[\hat{J}^{-}b^{+}+\hat{J}^{+}b]\eqa where here
the annihilation operator $b^{+}$ describes the cavity mode and
$g$ is the coupling constant between the qubits and the cavity.
Since this is symmetric in the atoms, it is natural to use the angular momentum
 states to describe thetwo two-level atoms.
 The most interesting dynamics occurs in the $j=1$ subspace.
 In terms of the individual atomic levels, the three
states for $j=1$ are
\beq \ket{1}=\ket{e}_{1}\ket{e}_{2},~~
\ket{2}=\frac{1}{\sqrt{2}}(\ket{g}_{1}\ket{e}_{2}+\ket{e}_{1}\ket{g}_{2}),~~
\ket{3}=\ket{g}_{1}\ket{g}_{2}. \eeq
The fourth state, i.e. the $j=0$ subspace, is \beq
\ket{\Psi_{j=0}}=\frac{1}{\sqrt{2}}(\ket{g}_{1}\ket{e}_{2}-\ket{e}_{1}\ket{g}_{2}).
\eeq This latter ssubspace will not change under any of  the
transformations we perform, as will be seen  below when we analyze
the adiabatic elimination of the cavity mode.

\subsection{Homodyne Detection}\label{homo}

For simplicity in explanation of homodyne detection, let us now
consider a system with a single two-level atom and subject the
atom to homodyne detection. We assume that all of the fluorescence
of the atom is collected and turned into a beam.
Ignoring the vacuum fluctuations in the field, the annihilation
operator for this beam is $\sqrt{\gamma}\sigma$, normalized so
that the mean intensity $\gamma\an{\sigma\dg \sigma}$ is equal to
the number of photons per unit time in the beam. This beam then
enters one port of a 50:50 beam splitter, while a strong local
oscillator $\beta$ enters the other. To ensure that this local
oscillator has a fixed phase relationship with the driving laser
used in the measurement, it would be natural to utilize the same
coherent light field source  both as the driving laser and as the
local oscillator in the homodyne detection. This homodyne
detection arrangement is as shown in
 Fig.~\ref{diag22}.

Again ignoring vacuum fluctuations, the two field operators for
the light exiting the beam splitter, $b_{1}$ and $b_{2}$, are \beq
b_{k} = \sq{\sqrt{\gamma}\sigma -(-1)^{k}\beta}/\sqrt{2}. \eeq
When these two fields are detected, the two photocurrents produced
have means \beq \label{meanI} \bar{I}_{k} = \an{ |\beta|^{2} -
(-1)^{k}\ro{ \sqrt{\gamma}\beta\sigma\dg + \sqrt{\gamma}\sigma
\beta^{*}} + \gamma\sigma\dg\sigma}/2. \eeq The middle two terms
represent the interference between the system and the local
oscillator.

Equation (\ref{meanI}) gives only the mean photocurrent. In an
individual run of the experiment for a system, what is recorded is not
the mean photocurrent, but the instantaneous photocurrent. This
photocurrent will vary stochastically from one run to the next,
because of the irreducible randomness in the quantum measurement
process. This randomness is not just noise, however. It is correlated
with the evolution of the system and thus tells the experimenter
something about the state of the system. 
The stochastic evolution of the state of the system conditioned by
the measurement record is called a ``quantum trajectory''
\cite{Car93b}. Of course, the master equation is still obeyed on
average, so the set of possible quantum trajectories is called an
unravelling of the master equation \cite{Car93b}. It is the
conditioning of the system state by the photocurrent record that
allows feedback control 
 the system state 
at the quantum level.

The ideal limit of homodyne detection is when the local oscillator
amplitude goes to infinity, which in practical terms means
$|\beta|^2\gg \gamma $. In this limit, the rates of the
photodetections go to infinity, and thus each photodetector
produces 
a continuous photocurrent with white noise. For our purposes the
only relevant quantity, suitably normalized, is the difference
between the two
photocurrents 
\cite{Car93b,WisMil93a} \beq \label{homo1} I(t) =
\frac{I_{1}(t)-I_{2}(t)}{|\beta|} =
\sqrt{\gamma}\an{e^{-i\Phi}\sigma\dg + e^{i\Phi}\sigma}_{\rm c}(t)
+ \xi(t). \eeq

A number of aspects of \erf{homo1} need to be explained. First,
$\Phi = \arg\beta$, the phase of the local oscillator (defined relative to the
driving field). Here we set $\Phi=0$. Of course, all that really matters here is
the relationship between the driving phase and the local oscillator
phase, not the absolute phase of either. Second, the subscript c means conditioned and refers
to the fact that if one is making a homodyne measurement then this
yields information about the system. Hence, any system averages will
be conditioned on the previous photocurrent record. Third, the final
term $\xi(t)$ represents Gaussian white noise, so that
\beq
\xi(t)dt = dW(t),
\eeq
an infinitesimal Wiener increment defined by \cite{Gar85}
\bqa
[dW(t)]^2=dt , \label{ito1}\\
{\rm E}[dW(t)]=0 . \label{dW0}
\eqa

\subsection{Adiabatic elimination of the cavity mode }

Let us now come back to the two-qubit system. The cavity mode is
uninteresting as for the high levels of damping its behaviour is
slaved to the driving, so it will then be adiabatically eliminated
\cite {thesis1,WisMil93c,Car2000} in our first calculation,
resulting in a master equation followed by the density operator
$\rho$, where $\rho$ only includes the two qubits.

The complete master equation describing the system pumped by an
unmodulated driving laser, including the cavity mode, is described
by a density operator $\omega$ given as follows. \bqa\label{old}
\dot{\omega} &=&
-\frac{1}{2}\alpha[b-b^{\dg},\omega]-i\frac{g}{2}[J^{-}b^{+}+J^{+}b,\omega]\nn\\
&&+\gamma_1{\cal D}\left[\sigma_1\right]\omega+\gamma_2{\cal
D}\left[\sigma_2\right]\omega+\gamma_{p}{\cal
D}\left[b\right]\omega. \eqa Here the annihilation operators $b$
describe the cavity mode. $\gamma_1, \gamma_2$ and $\gamma_p$ are
the coefficients of damping for the two qubits and $b$ mode
respectively. $\cal D$ is a superoperator defined as ${\cal
D}={\cal D}[A]B \equiv ABA\dg - \{A\dg A,B\}/2$ for irreversible
evolution. The cavity mode is heavily pumped and damped.

The adiabatic elimination of cavity mode is done first by
displacing the density operator $\omega$, and the master equation
describing its evolution, to zero in the $b$ mode. We assume that
$\gamma_{p}$ is sufficiently large that the state stays close to
an equilibrium coherent state with amplitude \beq
\beta={\alpha}/{\gamma_p}. \eeq The displacement operator \beq
D_{b}=e^{{\beta}(b^{+}-b)} \eeq is used to carry out this
transformation. The new density operator is $\nu=D_b(-\beta)\omega  D_b(\beta)$. Applying this operation to
the original master equation in Eq. (\ref {old}) gives the new
master equation for $\nu$ in which the $b$ mode is of zero average
amplitude. This is
\beq\label{v} \dot{\nu}={\cal
L}\nu-i\frac{g}{2}[(J^{+}b+{J}^{-}b^{+}),\nu]+{\gamma_p}{\cal
D}\left[b\right]\nu \eeq in which all the terms involving only the
qubits in the superoperator ${\cal L}$, defined as
\bqa {\cal
L}{\nu}=-i\frac{g\alpha}{2{\gamma_p}}[(J^{+}+J^{-}),\nu]+\gamma_1{\cal
D}\left[\sigma_1\right]\nu+\gamma_2{\cal
D}\left[\sigma_2\right]\nu \eqa

Since the amplitude of mode $b$ is small, a partial expansion of
the density matrix $\nu$ in terms of the $b$ mode number states
need only be carried out to small photon numbers. Therefore \bqa
\nu&=&\rho_0\ket{0}\bra{0}+(\rho_1\ket{1}\bra{0}+{\rm
H.c.})+\rho_2\ket{1}\bra{1}\nn\\
&&+({\rho_2}'\ket{2}\bra{0}+{\rm H. c.})+{\rm O}{(\lambda^3)},
\eqa where $\lambda={g}/{\gamma_p}$ is a very small number and
$\ket{0}$ is the cavity vacuum state. This is substituted into the
master equation Eq.(\ref{v}) which is expanded and terms
multiplying equal sets of $b$ mode number projectors are gathered
together to get a set of four equations. Terms of greater then
second order are neglected. These equations are: \bqa\label{four}
\dot{\rho_0}&=&{\cal L}\rho_0-i\frac{g}{2}[J^{+}\rho_1-\rho_1^{+}J^{-}]+{\gamma_p}\rho_2\nn\\
\dot{\rho_1}&=&{\cal
L}\rho_1-i\frac{g}{2}[J^{+}\rho_0+\sqrt{2}J^{+2}{\rho_2}'-J^{-2}{\rho_2}']
-\frac{\gamma_p}{2}\rho_1\nn\\
\dot{\rho_2}&=&{\cal L}\rho_2-i\frac{g}{2}[J^{+}\rho_1^{+}-{\rho_1}J^{+2}]-{\gamma_p}\rho_2\nn\\
\dot{{\rho_2}'}&=&{\cal
L}{\rho_2}'-i\frac{g}{2}[\sqrt{2}{J^{-}}\rho_1]-{\gamma_p}{\rho_2}'
\eqa Now we make the assumption that both $\dot{\rho}_1=0$ and
$\dot{{\rho_2}'}=0$ so that by using the second and fourth of
Eq.({\ref{four}}), the values of $\rho_1$ and ${\rho}_2'$ are
found to be \bqa
\rho_1&=&\frac{-ig}{\gamma_p}\left[J^{-}\rho_0-J^{-}\rho_2\right],\nn\\
{\rho_2}'&=&\frac{ig}{\sqrt{2}{\gamma_p}}J^{-}\rho_1, \eqa where
$\rho_1={\rm O}(\frac{g}{\gamma_p})$ and ${\rho_2}={\rm
O}(\frac{g^2}{\gamma_p})$. These are then substituted into the
first and third equation of set Eq.(\ref{four}) which become \bqa
\dot{\rho_0}&=&{\cal L}\dot{\rho_0}-\frac{g^2}{2\gamma_p}\big[J^{+}J\rho_0+
\rho_{0}J^{+}J^{-}-J^{+}J^{-}\rho_2\nn\\&&-\rho_{2}J^{+}J^{-}\big]+\gamma_p{\rho_2},\nn\\
 \dot{\rho_2}&=&{\cal L}\rho_2+\frac{g^2}{\gamma_p}\left[J^{-}{\rho_0}J^{+}-J^{-}\rho_2J^{+}\right]-\gamma_p{\rho_2}.
\eqa Adding these two equations together and noting that
${\rho_2}={\rm O}(\frac{g^2}{\gamma_p})$, then neglecting the
$\rho_2$ terms gives for the final master equation of the system.
\beq \dot{\rho}={\cal L}{\rho}+\frac{g^2}{\gamma_p}{\cal
D}\left[J^{-}\right]{\rho}, \eeq The coefficient
$\frac{g^2}{\gamma_p}$ describes the strength of collective
damping of the two-qubit and the cavity $b$ mode and will be
called $\gamma$.

The term ${\cal L}\rho$ is expanded to give
the master equation as \beq
\dot{\rho}=-i\frac{g\alpha}{2{\gamma_p}}[(J^{+}+J^{-}),\rho]+\gamma_1{\cal
D}\left[\sigma_1\right]\rho +\gamma_2{\cal
D}\left[\sigma_2\right]\rho+ \gamma{\cal D}\left[J^{-}\right]\rho
\eeq  This is the super-fluorescence master equation \cite{edit}.
The approximation that the timescale imposed by collective decay
rate $\gamma_p$ is greater than the timescale of the two qubits
evolution imposed by single decay rate $\gamma_{1},{\gamma_2}$,
that is $\gamma_1, \gamma_2{\ll}\gamma$, results in the following
master equation \beq
\dot{\rho}=-i\frac{g\alpha}{2{\gamma_p}}[(J^{+}+J^{-}),\rho]+\gamma{\cal
D}\left[J^{-}\right]\rho. \eeq

Following the method of Ref.\cite{WisMil93a}, the stochastic
master equation (SME) conditioned on homodyne measurement of the
output of cavity is \bqa d\rho_{\rm c} &=& -i[H, \rho_{\rm c}(t)]
+dt\gamma{\cal D}[J^{-}]\rho_{\rm c}(t)\nn\\&& +
\sqrt{\gamma}dW(t){\cal H}[J^{-}] \rho_{\rm c}. \eqa The homodyne
photocurrent, normalized so that the deterministic part does not
depend on the efficiency,is \beq \label{homo33} I(t) =
\sqrt{\gamma}\an{J_{x}}_{\rm c}(t) + \xi(t)/\sqrt{\eta}. \eeq

\subsection{Dynamics with Feedback}

If we now add dynamics with feedback from feedback Hamiltonian
$H_{\rm fb}=I(t)F$, where
$I(t)={\langle}{J^{+}+J^{-}}{\rangle}(t)+{\xi(t)}/{\sqrt{\eta}}$,
we can get the SME \bqa
\label{master2} d\rho_{\rm c} &=&  dt\gamma{\cal
D}[J^{-}]\rho_{\rm c} - idt [H_{\alpha},\rho_{\rm c}] {-}
idt[F,-iJ^{-}\rho_{\rm c} + i\rho_{\rm c}J^{+}]\nn\\
&&+dt\frac{1}{\gamma}{\cal D}[F]\rho_{\rm c} {+} dW(t){\cal
H}[-i\sqrt{\gamma}J^{-}-i{\lambda}J_{x}]\rho_{\rm c}. \eqa Here
$H_{\alpha}={\alpha}J_x$,~~$F={\lambda}J_x$, $J_x=J^{-}+J^{+}$ and
$\alpha, \lambda$ are driving and feedback amplitude respectively.
This corresponds to having a feedback-modulated driving laser.
This is also an \ito stochastic equation, which means that the
ensemble average master equation can be found simply by dropping
the stochastic terms.

Therefore from the above master equation in $j=1$ subspace, we can
write down the equation of motion for the components of the
$3\times3$ density matrix of the state of the system, taking into
account that $Tr(\hat{\rho})=1$ and that $\rho$ is Hermitian. In
order to get the steady state solution of the above master
equation Eq.(\ref{master2}), we define \bqa
x_{ij}=\bra{i}J_x\ket{j}+\bra{j}J_x\ket{i},\nn\\
y_{ij}=\bra{i}J_y\ket{j}+\bra{j}J_y\ket{i},\nn\\
z_{ij}=\bra{i}J_z\ket{j}+\bra{j}J_z\ket{i}.
\eqa
Then the differential equations for $x_{ij},y_{ij},z_{ij}$ are found to be
\bqa
\dot{x}_{12}&=&-2\gamma{x}_{12}-2\lambda+\frac{\lambda^2}{\gamma}{x}_{23}+\sqrt{2}\alpha{y}_{13},\nn\\
\dot{x}_{13}&=&(-\gamma-\frac{\lambda^2}{\gamma}-2\lambda){x}_{13}-\sqrt{2}\alpha{y}_{12}+\sqrt{2}{\alpha}{y}_{23},\nn\\
\dot{x}_{23}&=&(2\gamma+\frac{\lambda^2}{\gamma}){x}_{12}-(\gamma^2+\lambda^2){x}_{23}+\sqrt{2}\alpha{y}_{13},\nn\\
\dot{y}_{12}&=&-\alpha\sqrt{2}{x}_{13}+(2\gamma+6\lambda+\frac{5\lambda^2}{\gamma}){y}_{12}-2\sqrt{2}\alpha{y}_{13},\nn\\
\dot{y}_{13}&=&-\sqrt{2}\alpha{x}_{12}+\sqrt{2}\alpha{x}_{23}+(\gamma+2\lambda+\frac{2\lambda^2}{\gamma}){y}_{13},\nn\\
\dot{y}_{23}&=&\sqrt{2}\alpha{x}_{13}+(-2\gamma-\frac{3\lambda^2}{\gamma}-6\lambda){y}_{12}\nn\\&&+(\gamma+4\lambda+
\frac{5\lambda^2}{\gamma}){y}_{23}+2\sqrt{2}\alpha{z}_{12}-2\sqrt{2}\alpha{z}_{13},\nn\\
\dot{z}_{12}&=&(-2\lambda-\frac{3\lambda^2}{\gamma}){x}_{13}-2\sqrt{2}\alpha{y}_{12}+\alpha\sqrt{2}{y}_{23}-
\frac{8\gamma}{3}{z}_{12}\nn\\&&+(\frac{2\gamma}{3}-\frac{4\lambda}{3}+\frac{2\lambda^2}{\gamma}){z}_{13}-\frac{2\gamma}{3}
-\frac{4\lambda}{3},\nn\\
\dot{z}_{13}&=&2\lambda{x}_{13}-\sqrt{2}\alpha{y}_{12}-\sqrt{2}\alpha{y}_{23}+
(\frac{2\gamma}{3}+\frac{4\lambda}{3}){z}_{12}\nn\\&&-(\frac{4\gamma}{3}+\frac{8\lambda}{3}+\frac{2\lambda^2}{\gamma}){z}_{13}
-\frac{4\gamma}{3}+\frac{8\lambda}{3}. \eqa Here we have ignored
$z_{23}$ since we require only $8$ parameters. The steady-state
solutions are \bqa\label{steady1}
x_{12}=\frac{A}{S},~~x_{13}=\frac{B}{S},~~x_{23}=\frac{C}{S},~~y_{12}=\frac{D}{S}\nn \\
y_{13}=\frac{E}{S},~~y_{23}=\frac{F}{S},~~z_{12}=\frac{G}{S},~~z_{13}=\frac{H}{S}.
\eqa Here \bqa
S&=&24{\alpha}^4{\gamma}^4+2{\gamma}^8+26{\gamma}^7{\lambda}+
143{\gamma}^6{\lambda}^2+432{\gamma}^5{\lambda}^3\nn\\&&
+789{\gamma}^4{\lambda}^4+926{\gamma}^3{\lambda}^5+726{\gamma}^2{\lambda}^6
 +352{\gamma}{\lambda}^7+96{\lambda}^8\nn\\&&
+2{\alpha}^2{\gamma}^2\left(4{\gamma}^4+22{\gamma}^3{\lambda}+
65{\gamma}^2{\lambda}^2\right)+116{\gamma}{\lambda}^3+60{\lambda}^4,\nn\\
\eqa and \bqa
A&=&0 \nn \\
B&=&2\gamma(\gamma+2\lambda)(4\alpha^2\gamma^2(\gamma^2+3\gamma\lambda+\lambda^2)+ \nn \\
&&\lambda^2(2\gamma^4+14\gamma^3\lambda+33\gamma^2\lambda^2+32\gamma\lambda^3+16\lambda^4)) \nn \\
C&=&0 \nn \\
D&=&2\sqrt{2}\alpha\gamma^2(\gamma+2\lambda)(4\alpha^2\gamma^2+\lambda^2(5\gamma^2+
20\gamma\lambda+16\lambda^2)) \nn \\
E&=&0 \nn \\
F&=&2\sqrt{2}\alpha\gamma^2(\gamma+2\lambda)(4\alpha^2\gamma^2+2\gamma^4+14\gamma^3\lambda+ \nn\\
&&37\gamma^2\lambda^2+
44\gamma\lambda^3+44\gamma\lambda^3+16\lambda^4) \nn \\
G&=&\gamma(\gamma+2\lambda)(4\alpha^2\gamma^2(\gamma^2+3\gamma\lambda+\lambda^2)+ \nn \\
&&\lambda^2(2\gamma^4+14\gamma^3\lambda+33\gamma^2\lambda^2+32\gamma\lambda^3+16\lambda^4)) \nn \\
H&=&\gamma(\gamma+2\lambda)^3(4\alpha^2\lambda^2+2\gamma^4+ \nn \\
&&14\gamma^3\lambda+33\gamma^2\lambda^2+32\gamma\lambda^3+16\lambda^4).
\eqa

\section{Entanglement and purity}\label{measure}

Now that we have the steady-state solution of the master equation,
we can determine if feedback-modulated driving can increase the
steady-state entanglement when the collective decay is considered.
Let us now specify the measures which we will be using to
characterize the degree of entanglement of a state. As we
mentioned before, there are several measures of entanglement. In
this paper since we have two qubits system and we choose the
concurrence \cite{Wootters} as a measure for it. For a mixed state
represented by the density matrix $\rho$, the "spin-flipped"
density operator, which was introduced by Wootters
{~\cite{Wootters}}, is given by: \beq
\tilde{\rho}=\left({\sigma}_{y}\otimes{\sigma}_{y}\right)\overline{\rho}
\left({\sigma}_{y}\otimes{\sigma}_{y}\right), \eeq where the bar
of $\overline{\rho}$ denotes complex conjugate of ${\rho}$ in the
basis of $\left\{\ket{gg},\ket{ge},\ket{eg},\ket{ee}\right\}$, and
${\sigma}_{y}$ is the usual Pauli matrix given by \bqa
{\sigma}_{y} &=&\left(\begin{array}{cc}
0&-i\\
i&0\\
\end{array}\right) .\eqa  ~~~In order to work out the concurrence,
we need to determine the square root of the eigenvalues
$\lambda_1,\lambda_2,\lambda_3,\lambda_4$ of the matrix
${\rho}\tilde{\rho}$ and sort them in decreasing order, i.e.,
$\lambda_1{\geq}\lambda_2{\geq}\lambda_3{\geq}\lambda_4$. It can
be shown that all these eigenvalues are real and non-negative. The
concurrence $C$ of the density matrix ${\rho}$ is defined as \beq
C\left({\rho}\right)={\rm
max}\left(\sqrt{\lambda_1}-\sqrt{\lambda_2}-\sqrt{\lambda_3}-\sqrt{\lambda_4},0\right)
\eeq  The range of concurrence $C$ is $0$ to $1$. When $C$ is
nonzero the state is entangled. The maximum entanglement is when
$C=1$.

\begin{figure}
\centerline{\scalebox{0.4}{\includegraphics{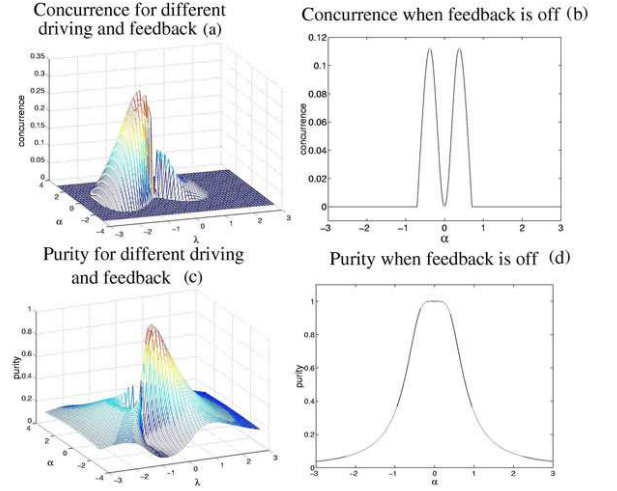}}}
\caption{Plot of concurrence and purity vs driving and feedback
amplitude.} \label{ent11}
\end{figure}

Fig. \ref{ent11} is the plot of concurrence and purity vs driving and
feedback amplitude. As shown in Fig.\ref{ent11} a2, we can get
certain amount of entanglement in the steady-state of a
unmodulated driving system \cite {thesis1}. In this case the
maximum concurrence is about 0.11, with appropriate choice of the
driving amplitude $\alpha=0.38\gamma^{-1}$. The steady state is
independent of the initial state as long as it is in the symmetric
subspace, such as $\ket{g}\bra{g}$.  The coherent evolution alone
is not able to produce any entanglement for an initially
unentangled state, as it only consists of single qubit rotation
with no coupling between two qubits. Therefore the steady-state
entanglement is due to the common cooperative decoherence coupling
to the cavity environment acting together with the coherent
evolution \cite{thesis1}.

After including feedback onto the amplitude of the driving on the
atom, proportional to the homodyne photocurrent, we see that
feedback is remarkable as the steady-state concurrence has been
improved from $0.11$ to $0.31$ as shown in Fig.\ref{ent11} a1,
with appropriate choice of driving amplitude
$\alpha=\pm{0.4}{\gamma^{-1}}$ and feedback amplitude
$\beta=-0.8\gamma^{-1}$.

The gain of the steady state entanglement comes at the price of a
loss of purity, as shown in Fig.~\ref{ent11} (b).  There are a
number of measures that can be used for the degree of purity, for
example, the von Neumann entropy given by $S=-\rm
Tr[{\rho}\ln{\rho}]$, and the trace sqared of the density matrix.  In
this paper we choose the measure of purity given by \beq
r^2=\frac{4}{3}(\rm Tr[{{\rho}}^2]-\frac{1}{4}). \eeq From the
above equations, we find that \bqa
r^2&=&\frac{1}{9}+\frac{8}{9}\big[x_{12}^2+x_{13}^2+x_{23}^2+y_{12}^2+y_{13}^2+x_{23}^2+z_{12}^2\nn\\
&&+\frac{1}{3}(2z_{13}-z_{12})^2\big]. \eqa The minimum purity in
$3\times3$ subspace is obviously $\frac{1}{9}$. Note that Eq(36)
measure is linear in $\rm Tr[{{\rho}}^2]$ with minimum $0$ and
maximum $1$. The minimum of $0$ is only attainable in $4\times4$
Hilbert space.

\begin{figure}
\centerline{\scalebox{.5}{\includegraphics{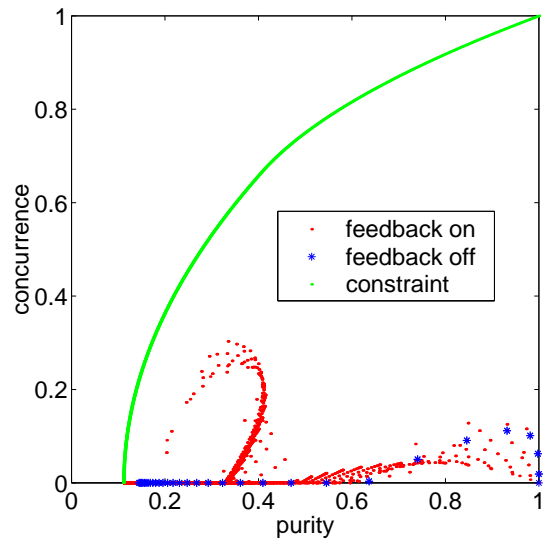}}}
\caption[Plot of concurrence vs purity]{Plot of concurrence vs
purity. The dots representing the case of feedback-modulated
driving laser with two adjustable parameters $\alpha$ and
$\lambda$, and the stars the case of unmodulated driving only with
one adjustable parameter $\alpha$. The continuous diagonal curve
tands for the maximally-entangled mixed states, states with the
maximal amount of entanglement for a given degree of purity  }
\label{con}
\end{figure}

To gain further insight into the purity of the steady-state when
it is  entangled, we may look at the position of the
steady-state located in the concurrence-purity plane. We begin by
choosing a series of driving and feedback amplitudes, and
determine their corresponding purity and entanglement. In
Fig.\ref{con}, we display these results with dots, representing
the case of feedback-modulated driving, and stars,representing the
case of unmodulated driving. Obviously, the feedback mechanism
leads to a noticeable increase of entanglement, though it results
in a less pure state. The continuous curve in Fig.\ref{con}
represents the maximally-entangled mixed states, states with the
maximal amount of entanglement for a given degree of purity, or in
other words, states with the minimum purity for a given
concurrence. The concurrence and the purity of a maximally
entangled mixed state satisfy the following equation \cite{munro}
\beq r^2={1-\frac{3}{4}[4g(2-3g)-C^2]}, \eeq where
$$
g=\left\{
\begin{array}{cc}
{C}/{2}&~~~~{\rm for}~~C>{2}/{3}\\
{1}/{3}&~~~~{\rm for}~~C<{2}/{3}
\end{array}
\right.
$$
In reference \cite{munro}, the degree of the mixture of a state is
defined by linear entropy $S_L$. The purity $r^2$ defined in this
paper is related to $S_L$ by $r^2=1-S_L$.

\section{Q function and density matrix}\label{Q and}

\begin{figure}
\centerline{\scalebox{.45}{\includegraphics{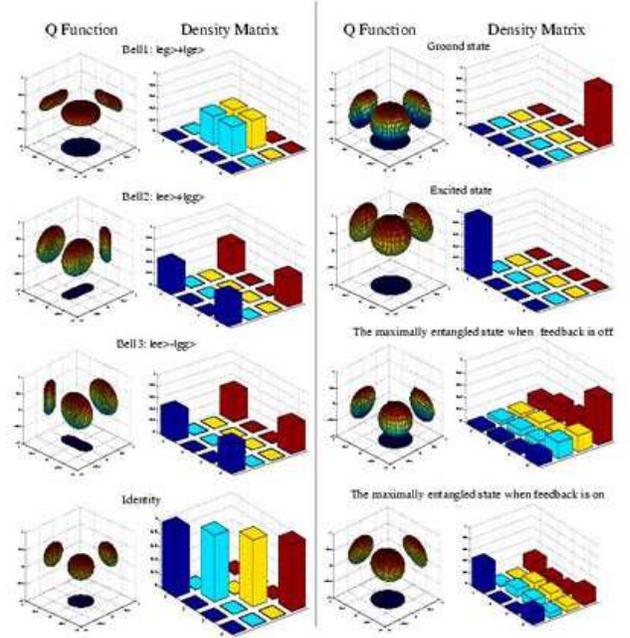}}}
\caption[Plot of the $Q$ function and density matrix]{Plots of the
$Q$ function and its projections on each of three planes, and the
absolute values of matrix elements of the density matrix are shown
for each of eight different states, as indicated in the figure.
The magnitude of the $Q$ function in a particular direction is
represented by the distance measured from the center of the $Q$
function.The diagonal elements of the density matrix are
$\ket{ee}\bra{ee},
\ket{eg}\bra{eg},\ket{ge}\bra{ge},\ket{gg}\bra{gg}$ from left to
right respectively.}\label{QD1}
\end{figure}

~~~~To gain further knowledge about the nature of the steady-state
when it is maximally entangled, we look at the density matrix and
$Q$ function. Since the density matrix elements are complex
numbers, we plot in Fig.\ref {QD1} the modulus of the matrix
elements. We use the unentangled basis states \bqa
\ket{\psi_1}&=&\ket{g}_1\ket{g}_2,\\ \nn
\ket{\psi_2}&=&\ket{g}_1\ket{e}_2,\\ \nn
\ket{\psi_3}&=&\ket{e}_1\ket{g}_2,\\ \nn
\ket{\psi_4}&=&\ket{e}_1\ket{e}_2. \eqa  because a separable basis
is needed to discuss entanglement.

To define the $Q$ function we need atomic coherent state
$\ket{\zeta}$ ~\cite{arecchi,radcliffe} \beq\label{coherent}
\ket{\zeta}=(1+{|(\zeta)}|^2)^{-j}\exp[{\zeta}\hat{J^{+}}]\ket{j,-j}.
\eeq The atomic coherent state is the closest quantum mechanical
states to a classical description of a spin system.

The $Q$ function is a positive distribution function, which is defined as
\bqa\label{Q}
Q(\theta,\phi)&=&{\langle}j,\theta,\phi|\rho|j,\theta,\phi{\rangle} \nn \\
|j,\theta,\phi{\rangle}&=&\sqrt{\frac{(2j)!}{(j+m)!(j-m)!}}\nn\\&&\sum_{m=-j}^{j}\left[\cos^{j+m}(\frac{\theta}{2})\sin^{j-m}
(\frac{\theta}{2})e^{-im{\phi}}\right]|j,m{\rangle},\nn\\ \eqa
where the state $|j,\theta,\phi{\rangle}$ is the spherical
representation of the coherent state  $\ket{\zeta}$ where $\theta$
and $\phi$ are the standard spherical polar coordinates defined by
$\zeta=\rm arctan{\frac{\theta}{2}}{e^{-i\phi}}$. The magnitude of
the $Q$ function in a particular direction ($\theta$ and $\phi$)
is represented  in Fig.\ref {QD1} by the distance measured from
the origin. Since the $Q$ function is the projection of the
density matrix into a coherent state , the $Q$ function does not
give more information than the density matrix. However, the
advantage of the $Q$ function is that it gives a more intuitive
view of where the state is located.

We plot the steady-state $Q$ function and density matrix in
Fig.\ref {QD1}.  There are a number of points that need
explanation. First, when there is only the unmodulated driving
laser shining on the system, the steady state having the most
entanglement is mainly confined to the ground state $Q$ function
and the upper state is almost unpopulated. However, when the
driving amplitude is modified by feedback, the upper state becomes
very well populated and the entanglement is greatly increased.
This is not surprising, as more population in the upper state
enhances nonlinearity both in the decoherence coupling and
*coherent* evolution of the two-qubit system.

Second, in the particular separable $4\times4$ basis , the
off-diagonal elements of the density matrix represent coherence
and the presence of coherence is a necessary condition for the
creation of entanglement. To illustrate this, we plot three Bell
states which are the maximally entangled states in Fig.\ref {QD1}.
We see that the off-diagonal terms are present in these maximally
entangled states. In contrast, we also plot three non-entangled
states, the identity, the ground state and the excited state. We
see that none of the off-diagonal elements appear.

Third, when there is only the unmodulated driving laser shining on
the system, the density matrix of the most entangled steady-state
looks most similar to that of  superposition of $\ket{gg}$ and
some of a Bell state $\frac{1}{\sqrt{2}}(\ket{eg}+\ket{ge})$,
while when the feedback modulation is switched on, the
steady-state density matrix looks more like a mixture of a
non-maximally entangled state $a\ket{eg}+b\ket{ge}$ and
$\ket{ee}+\ket{gg}$.

\section{Discussion}\label{discussion}

To summarize, entanglement between two qubits can be created
dynamically by driving and coupling them to a heavily damped
cavity mode. When there is only the unmodulated driving laser
shining on the system, the maximum steady-state concurrence (a
measure of entanglement) is 0.11 \cite{thesis1}. In this paper we
have re-derived these results and constructed a scheme to increase
the steady state entanglement by using homodyne-mediated feedback,
in which the driving laser is modulated by the homodyne
photocurrent derived from the cavity output.

An analytical form for the steady-state solution of the master
equation with feedback was derived and was used to show that the
amount of the maximum steady-state concurrence has been increased
from 0.11 to 0.31. The properties of the entangled state were also
studied through the discussion of the $Q$ function and the density
matrix. The important point here is that the feedback scheme can
lift the steady-state from the ground state towards the excited
state. Indeed with such a feedback scheme the most entangled state
much closer to the Bell states, which are maximally entangled
states. The details about how the feedback mechanism dynamically
changes the position of the steady-state when it is maximally
entangled are still largely unknown and require further
investigation. Open questions such as ``can the most entangled
steady-state be realized experimentally?", ``what happens if
individual decay of each atom cannot be ignored in the
calculation", ``how does one increase both entanglement and the
purity of a mixed state", ``to what extent can the system be
treated semi-classically" and ``what happens for a higher number
of atoms" are subjects for future exploration.

\section{Acknowledgments}

Jin Wang would like to acknowledge stimulating discussions with
Dr. Herman Batelaan, Dr. Hong Gao, Dr. Mikhail Frolov, Andrei Y.
Istomin, Amiran Khuskivadze, and Dr. Anthony Starace as well
as support from the Nebraska Research Initiative on Quantum
Information.  Jin Wang would also like to thank Jeremy Podany for
proofreading. H. M. Wiseman and G. J. Milburn would like to
acknowledge the support of the Australian Research Council Special
and the State of Queensland.


\end{document}